\documentstyle[12pt,epsf]{article}
\begin{document}

\begin{titlepage}

\begin{flushright}
EPHOU-98-012 \\ 
October 1998 \ \ \ 
\end{flushright}

\vspace{15mm}

\begin{center}
{\large BPS Analysis of the Charged Soliton Solutions \\ 
        of D-brane Worldvolume Theory \\
        from the Viewpoint of Target-space Supersymmetry }  
\end{center}

\vspace{5mm}

\renewcommand{\thefootnote}{\fnsymbol{footnote}}

\begin{center}
Toshiya Suzuki \footnote[1]{tsuzuki@particle.sci.hokudai.ac.jp} 
\end{center}

\begin{center}
{\it Department of Physics, Hokkaido University \\
     Sapporo 060-0810, Japan } 
\end{center}

\vspace{5mm}

\begin{center}
{\bf Abstract}
\end{center}

We investigate BPS properties of the charged soliton solutions of D-brane worldvolume theory, which is described by the supersymmetric Dirac-Born-Infeld action, by means of the $N=2$ target-space supersymmetry algebra. Our results agree with those obtained previously. We also extend our BPS analysis to the case where axion background exists. 

\end{titlepage}

\setcounter{footnote}{0}

D-branes play an important role in non-perturbative physics of superstring \cite{P,W}. Their dynamics are described by the supersymmetric Dirac-Born-Infeld (DBI) action \cite{APS1,APS2,CGNW,CGNSW,BT}, which have the worldvolume reparametrization invariance, U(1) gauge symmetry, $\kappa$-symmetry ( all of them are local symmetries ) and the global $N=2$ target-space supersymmetry as well. The algebra of the Noether charges associated with the global supersymmetry was calculated by making use of the Poisson bracket and explicit form of the central charges, which are present in the algebra, was given \cite{H,HK1,HK2} \footnote{ Before \cite{H,HK1,HK2}, the existence of the central charge in the supersymmetry algebra was shown for the M-5-brane theory \cite{ST} .} .

The DBI theory is known to admit soliton solutions \cite{CM,G,GGT}, some of which carry the electric and/or magnetic U(1) charges. Their BPS properties that they are invariant under portion of supersymmetry transformation have been discussed by several authors \cite{CM,G,GGT,LPT}. 

In this letter, we consider this issue from the viewpoint of the target-space supersymmetry by making use of the result of \cite{H,HK1,HK2}. Our results agree with those obtained previously.  

Our analysis will be performed only in the type IIB D-brane theory. For type IIB theory, we can turn on axion background field \cite{APPS}. We calculate the central charges which include the effects of the background, and use these results in our BPS analysis of charged soliton solutions.  

Let us briefly review the type IIB super D-brane actions in the flat background \cite{APS1,APS2,HK1}. They are described by the target-space vector $X^{m}$ , two Majorana-Weyl spinors $\theta^{\alpha}_{A}$ ($A=1,2$) with the same chirality and the world-volume U(1) gauge field $A_{\mu}$ ($\mu=0,1,...,p$) . The action decomposes into two parts 
\begin{equation}
S = \int d^{p+1} \sigma {\cal L}_{DBI} + \int L_{WZ} .
\label{action}
\end{equation}
The first term is the supersymmetrized DBI action 
\begin{equation}
{\cal L}_{DBI} = - e^{-\phi} \sqrt{-det(G_{\mu \nu}+{\cal F}_{\mu \nu})} , 
\label{DBI}
\end{equation}
where $\phi$ is constant dilaton background field. $G_{\mu \nu}$ and ${\cal F}_{\mu \nu}$ are defined as 
\begin{equation}
G_{\mu \nu} = \Pi_{\mu}^{m} \Pi_{\nu m}
\label{Gmn}
\end{equation}
\begin{equation}
{\cal F}_{\mu \nu} = F_{\mu \nu} - \{ \bar{\theta} \tau_{3} \Gamma_{m} \partial_{\mu} \theta ( \Pi_{\nu}^{m} + \frac{1}{2} \bar{\theta} \Gamma^{m} \partial_{\nu} \theta ) - ( \mu \leftrightarrow \nu ) \}
\label{Fmn}
\end{equation}
where $ \Pi_{\mu}^{m} = \partial_{\mu} X^{m} - \bar{\theta} \Gamma^{m} \partial_{\mu} \theta $ , $ F_{\mu \nu} = \partial_{\mu} A_{\nu} - \partial_{\nu} A_{\mu} $ and $\tau$'s are the Pauli matrices which act on index $A=1,2$ .
The second term is the Wess-Zumino (WZ) term, the definition of which is given in \cite{APS1,APS2} through its derivative 
\begin{eqnarray}
dL_{WZ} & = & e^{-\phi} d \bar{\theta} S_{B} \tau_{1} d \theta e^{\cal F} \nonumber \\
S_{B}({\Pi \!\!\! /}) & = & \sum_{l=0} \tau_{3}^{l} \frac{{\Pi \!\!\! /}^{2l+1}}{(2l+1)!} ,
\label{WZ}
\end{eqnarray}
while the integral form is given in \cite{HK1} .

The global supersymmetry transformation laws are 
\begin{eqnarray}
\delta_{\epsilon} \theta & = & \epsilon \nonumber \\
\delta_{\epsilon} X^{m} & = & \bar{\epsilon} \Gamma^{m} \theta \nonumber \\
\delta A & = & \bar{\epsilon} \tau_{3} \Gamma_{m} \theta dX^{m} - \frac{1}{6} \bar{\epsilon} \tau_{3} {V \!\!\!\! /} \theta 
\label{susy}
\end{eqnarray}
where $ V^{m} = (\bar{\theta} \Gamma^{m} d \theta) + \tau_{3} (\bar{\theta} \tau_{3} \Gamma^{m} d \theta) $ . Under this transformation, the action is invariant up to total derivative. The corresponding Noether charge, which generates the transformation (\ref{susy}), is
\begin{eqnarray}
Q \epsilon & = & \int d^{p} \sigma ( p_{m} \delta_{\epsilon} X^{m} + \zeta \delta_{\epsilon} \theta + E^{a} \delta_{\epsilon} A_{a} ) \nonumber \\
 & & - e^{-\phi} \int d^{p} \sigma [q_{\epsilon}^{2}]_{\bf p} \nonumber \\
 & = & Q_{1} \epsilon + Q_{2} \epsilon .
\label{Q}
\end{eqnarray}
where $p_{m}$ , $\zeta$ and $E^{a}$ are the conjugate momenta of $X^{m}$ , $\theta$ and $A_{a}$ ($a=0,1,...,p-1$) . The second term arises from the variation of the WZ term 
\begin{equation}
\delta_{\epsilon} L_{WZ} = e^{-\phi} d [q_{\epsilon}^{2}] .
\label{Q_WZ}
\end{equation}
The explicit form of $ q_{\epsilon}^{2} $ is given in \cite{HK1}. In (\ref{Q}), $ [q_{\epsilon}^{2}]_{\bf p} $ denotes the coefficient of spatial p-form $ q_{\epsilon}^{2} $ . 

The Poisson bracket of (\ref{Q}) is calculated in \cite{H,HK1,HK2} , where it is found that the central charges appear in the algebra. For bosonic configuration ( $ \theta = 0 $ ) , their result is  
\begin{eqnarray}
\{ Q_{\alpha A} , Q_{\beta B} \} & = & -2 (C \Gamma_{m})_{\alpha \beta} \delta_{A B} \int d^{p} \sigma p^{m} \nonumber \\ 
 & & -2 (\tau_{3})_{A B} (C \Gamma_{m})_{\alpha \beta} \int d^{p} \sigma \partial_{a} X^{m} E^{a} \nonumber \\
 & & - (\tilde{\tau}_{J})_{A B} (C \Gamma_{M})_{\alpha \beta} \frac{2e^{-\phi}}{2^{5}} (1 + (-)^{\sigma_{M} + J}) \int d^{p} \sigma [U_{J}^{M}]_{\bf p} 
\label{algebra}
\end{eqnarray}
The third term, precise form of which is given in \cite{HK1}, comes from $ \{ Q_{1} , Q_{2} \} $ and $ \{ Q_{2} , Q_{1} \} $ ,while the first and second terms arise from $ \{ Q_{1} , Q_{1} \} $ . Note that $ Q_{2} $ includes no conjugate momentum, so $ \{ Q_{2} , Q_{2} \} $ has no contribution. For $p=1,3$ cases, the third terms are  
\begin{equation}
2 e^{-\phi} (C \Gamma_{m})_{\alpha \beta} (\tau_{1})_{A B} \int d \sigma \partial X^{m} = (C \Gamma_{m})_{\alpha \beta} (\tau_{1})_{A B} Z_{1}^{m} 
\label{Z1}
\end{equation}
and 
\begin{eqnarray}
 & & 2 e^{-\phi} (C \Gamma_{m})_{\alpha \beta} (\tau_{1})_{A B} \int d^{3} \sigma \epsilon^{a b c} \partial_{a} X^{m} \partial_{b} A_{c} \nonumber \\
 & + & \frac{2e^{-\phi}}{3!} (C \Gamma_{m_{1} m_{2} m_{3}})_{\alpha \beta} (i \tau_{2})_{A B} \int d^{3} \sigma \epsilon^{a b c} \partial_{a} X^{m_{1}} \partial_{b} X^{m_{2}} \partial_{c} X^{m_{3}}  \nonumber \\
 & = & (C \Gamma_{m})_{\alpha \beta} (\tau_{1})_{A B} Z_{1}^{m} + (C \Gamma_{m_{1} m_{2} m_{3}})_{\alpha \beta} (i \tau_{2})_{A B} Z_{2}^{m_{1} m_{2} m_{3}} 
\label{Z3}
\end{eqnarray}
respectively. Here we defined $Z_{1}^{m}$ and $Z_{2}^{m}$ by (\ref{Z1}) and (\ref{Z3}). Also we define $Z_{3}^{m}$ as 
\begin{equation}
Z_{3}^{m} = -2 \int d^{p} \sigma \partial_{a} X^{m} E^{a} . 
\label{univZ3}
\end{equation}

Usually the second term of (\ref{algebra}) and the second term of (\ref{Z3}) are total derivative terms, because of the Gauss law $ \partial_{a} E^{a} = 0 $ and the Bianchi identity $ \partial_{a} B^{a} = 0 $ where $ B^{a} = \epsilon^{a b c} \partial_{b} A_{c} $ . However, as shown in \cite{CM,G,GGT}, they can be $ \delta $ function valued quantities, so they represent electrically and/or magnetically charged configurations. In order to make the configurations BPS saturated, they must be accompanied by excitations of some scalar fields. In \cite{CM,G}, their BPS properties are shown for the linearized theory. The proof for the full non-linear theory is given in \cite{LPT}. Also, in \cite{GGT}, the Bogomol'nyi inequalities are derived by Hamiltonian analysis of the bosonic sector.

Now, we consider this issue, from the viewpoint of the target-space supersymmetry algebra (\ref{algebra}). We start with $ p = 1 $ case. For $ p = 1 $ , the electric BPS configuration is given \cite{DM} 
\begin{eqnarray}
E^{1} & = & -q \ \ \ \ \ \ \ \ for \ \ \sigma < 0 \nonumber \\
 & & 0 \ \ \ \ \ \ \ \ \ \ for \ \ \sigma > 0 \nonumber \\
X^{9} & = & - e^{\phi} q \sigma \ \ \ for \ \ \sigma < 0 \nonumber \\
 & & 0 \ \ \ \ \ \ \ \ \ \ for \ \ \sigma > 0 
\label{sol-p1}
\end{eqnarray}
and $ X^{i} = \sigma^{i} (i=0,1) $ . The Gauss law is modified to $ \partial_{1} E^{1} = q \delta (\sigma) $ . This configuration represents the 3-string junction \cite{S_junc} . 
\begin{center}
\begin{minipage}{70mm}
\epsfxsize=70mm \epsfbox {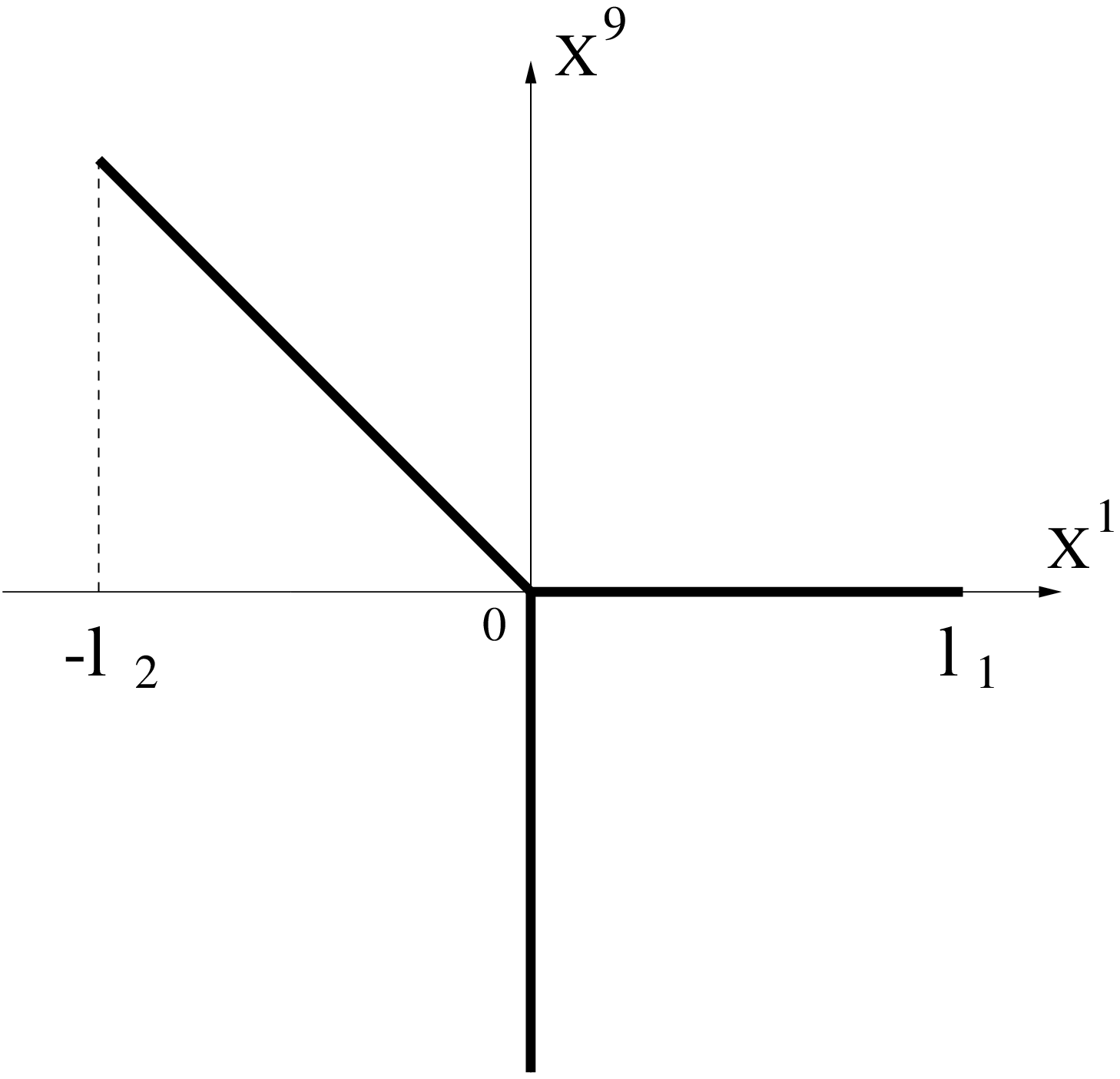}
\end{minipage} \\
The 3-string junction
\end{center}
Here, we assumed that the electric charge sits at the origin and that $X^{1}$ direction compactifies to $ - l_{1} \leq X^{1} \leq l_{2} $ . Also, we add fundamental string (F-string) along $X^{9}$ direction by hand, in order to interpret the electric charge on the D-string as the end-point of it \footnote{ This ``extra'' F-string was introduced in \cite{DM} , and its justification was discussed in \cite{Hash} . } . We will use $-X^{9}(0)$ to denote the length of the attached F-string. 

For this configuration, non-vanishing charges are  
\begin{eqnarray}
Z_{1}^{1} & = & 2 e^{-\phi} ( l_{1} + l_{2} ) \nonumber \\
Z_{3}^{9} & = & -2q ( - X^{9} (0) + e^{\phi} q l_{2} ) \nonumber \\
Z_{1}^{9} & = & -2q l_{2} \nonumber \\
Z_{3}^{1} & = & 2q l_{2}   
\label{cha-p1}
\end{eqnarray}
and $ P^{0} = \int d \sigma p^{0} $ . Note that $ X^{0} = \sigma^{0} $ , so $ P^{0} $ is the energy. Replacing the Poisson bracket with the anti-commutator ( divided by $i$ ) , we get the Bogomol'nyi inequality \footnote{ The charge conjugation matrix $C$ is defined as $ C = i \Gamma_{0} $ , so no factor $i$ emerges in the R.H.S. of the Bogomol'nyi inequality . }  
\begin{eqnarray}
0 & \leq & \{ 4(P^{0})^{2} - (Z_{1}^{1}+Z_{3}^{9})^{2} - (Z_{1}^{9}-Z_{3}^{1})^{2} \}^{8} \nonumber \\
 & & \{ 4(P^{0})^{2} - (Z_{1}^{1}-Z_{3}^{9})^{2} - (Z_{1}^{9}+Z_{3}^{1})^{2} \}^{8} . 
\label{ineq-p1-1}
\end{eqnarray}
From (\ref{cha-p1}), we see 
\begin{equation}
Z_{1}^{9} + Z_{3}^{1} = 0 
\label{cancel}
\end{equation}
so (\ref{ineq-p1-1}) reduces to 
\begin{equation}
2P^{0} \geq Z_{1}^{1}-Z_{3}^{9} .
\label{ineq-p1-2}
\end{equation} 
The bound is the same as (twice) the energy of the configuration (\ref{sol-p1}), so this is $\frac{1}{4}$ supersymmetric marginally BPS state.

To understand the meaning of the relation (\ref{cancel}), we return to the integral form  
\begin{eqnarray}
Z_{1}^{9} &=& 2 e^{-\phi} \int_{-l_{2}}^{0} d \sigma \partial X^{9} \nonumber \\
Z_{3}^{1} &=& -2 \int_{-l_{2}}^{0} d \sigma \partial X^{1} E^{1} . 
\label{cha-p1_int}
\end{eqnarray}  
From this, (\ref{cancel}) implies $ X^{9} = - e^{\phi} E^{1} \sigma $ , which is exactly given in (\ref{sol-p1}) . The slope $ - e^{\phi} E^{1} $ is necessary for the tensions of the constituent strings to balance, as shown in \cite{DM}. So we see that the tension-balance condition is needed for the solution to be marginal. 

Due to being marginally BPS state, the energy of the 3-string junction is the sum of the masses of 3 strings, that is, $(0,1)$-string, $(q,0)$-string and $(q,1)$-string \cite{Hash,RY} 
\begin{equation}
P^{0} = e^{-\phi} l_{1} + q (-X^{9}(0)) + T_{q,1} \frac{l_{2}}{\cos \alpha} 
\label{ene-p1}
\end{equation}
where $ T_{q,1} = \sqrt{e^{-2\phi}+q^{2}} $ and $ \cos \alpha = \frac{1}{\sqrt{1+q^{2}e^{2\phi}}} $ . 

In type IIB theory, we can turn on axion background field. It is known that, under this situation, the tension-formula of the $(q,g)$-string \footnote{ Usually, called $(p,q)$-string. } is changed to $SL(2,{\bf Z})$-covariant form \cite{S_pq} . So it is interesting to extend our BPS analysis to the case with non-vanishing axion background. 

We restrict our analysis to the theory with constant background, where the D-brane action is invariant under the same supersymmetry transformation as (\ref{susy}). In this case, the action is supplemented by a total derivative term \cite{APPS} 
\begin{eqnarray}
S = \int d \sigma {\cal L}_{DBI} + \int L_{WZ} - \int  C_{0} F 
\label{action-p1_da}
\end{eqnarray}
where the first and second terms are the same as (\ref{DBI}) and (\ref{WZ}) , and $C_{0}$ is axion. 

The supersymmetry algebra is supplemented by a $C_{0}$-dependent term, and then the form of the central charge is changed ( see Appendix A ) 
\begin{eqnarray}
Z_{1}^{m} & = & 2 e^{-\phi} \int d \sigma \partial X^{m} \nonumber \\
Z_{3}^{m} & = & -2 \int d \sigma \partial X^{m} E^{1} -2 C_{0} \int d \sigma \partial X^{m} .
\label{Z1_da}
\end{eqnarray}

The solution for this case is given \cite{DM} ( see also Appendix B ) 
\begin{eqnarray}
X^{9} & = & e^{\phi} ( -q + C_{0} ) \sigma \ \ \ \ \ for \ \ \ \sigma < 0 \nonumber \\
 & & e^{\phi} C_{0} \sigma \ \ \ \ \ \ \ \ \ \ \ \ \ \ \ for \ \ \ \sigma > 0 \nonumber \\
E^{1} & = & -q \ \ \ \ \ \ \ \ \ \ \ \ \ \ \ \ \ \ \ for \ \ \ \sigma < 0 \nonumber \\
 & & 0 \ \ \ \ \ \ \ \ \ \ \ \ \ \ \ \ \ \ \ \ \ for \ \ \ \sigma > 0 .
\label{sol-p1_da}
\end{eqnarray}
\begin{center}
\begin{minipage}{70mm}
\epsfxsize=70mm \epsfbox {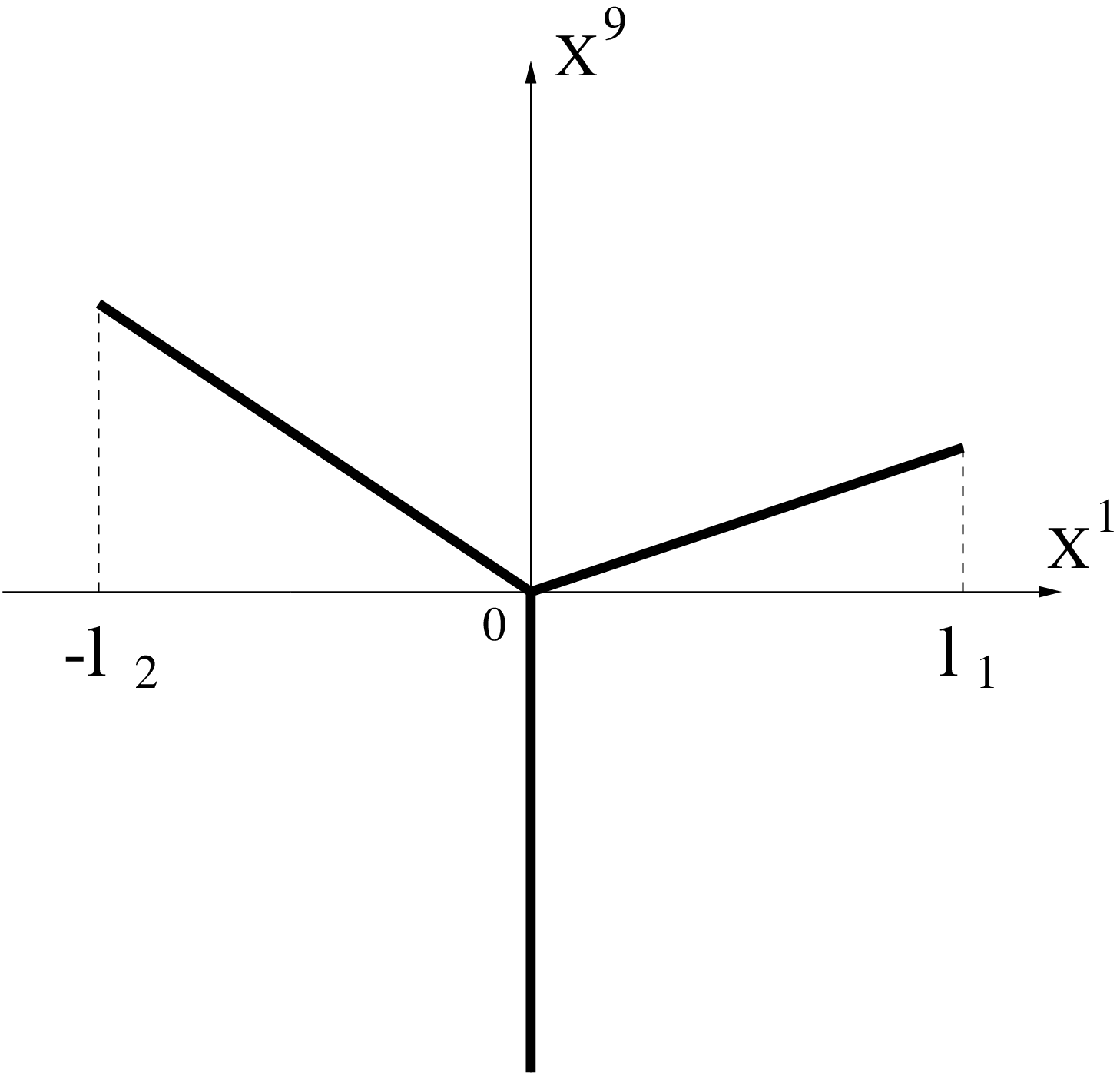}
\end{minipage} \\
The modified 3-string junction
\end{center}
From (\ref{Z1_da}) , (\ref{sol-p1_da}) 
\begin{eqnarray}
Z_{1}^{1} & = & 2 e^{-\phi} ( l_{1} + l_{2} ) \nonumber \\
Z_{3}^{9} & = & -2 e^{\phi} C_{0}^{2} l_{1} -2 e^{\phi} ( C_{0} - q )^{2} l_{2} + 2 q X^{9}(0) \nonumber \\
Z_{1}^{9} & = & 2 C_{0} ( l_{1} + l_{2} ) - 2 q l_{2} \nonumber \\
Z_{3}^{1} & = & 2 q l_{2} - 2 C_{0} ( l_{1} + l_{2} ) 
\label{cha-p1_da}
\end{eqnarray}
so we find $ Z_{1}^{9} + Z_{3}^{1} = 0 $ , then the Bogomol'nyi inequality reduces to $ 2 P^{0} \geq Z_{1}^{1} - Z_{3}^{9} $ again. This inequality is saturated by the configuration (\ref{sol-p1_da}), so this is $\frac{1}{4}$ supersymmetric marginally BPS configuration, and the energy is the sum of the masses of constituent strings
\begin{equation}
P^{0} = T_{0,1} \frac{l_{1}}{\cos \beta} + T_{q,0} (-X^{9}(0)) + T_{q,1} \frac{l_{2}}{\cos \alpha}
\label{ene-p1_da}
\end {equation}
where $ T_{0,1} = \sqrt{C_{0}^{2}+e^{-2\phi}} $ , $ T_{q,0} = q $ , $ T_{q,1} = \sqrt{(C_{0}-q)^{2}+e^{-2\phi}} $  and $ \cos \alpha = \frac{1}{\sqrt{1+e^{2\phi}(C_{0}-q)^{2}}} $ , $ \cos \beta = \frac{1}{\sqrt{1+e^{2\phi}C_{0}^{2}}} $ , which agree the Schwarz's tension formula in the presence of axion background \cite{S_pq} . 

Next, we turn to $ p = 3 $ case. This theory admits electrically and magnetically charged solution \cite{CM,G,GGT} , which is called BI Dyon. The BI Dyon solution is given as   
\begin{eqnarray}
E^{a} & = & \frac{q}{4 \pi r^{2}} \hat{r}^{a} \nonumber \\
B^{a} & = & \frac{g}{4 \pi r^{2}} \hat{r}^{a} \nonumber \\ 
X^{9} & = & - \frac{\sqrt{e^{2\phi}q^{2}+g^{2}}}{4 \pi r}
\label{sol-p3}
\end{eqnarray} 
and $ X^{\mu} = \sigma^{\mu} ( \mu = 0,1,2,3 ) $ , where $\hat{r}^{a}$ denotes the unit vector in the $r^{a}$ direction.  
\begin{center}
\begin{minipage}{70mm}
\epsfxsize=70mm \epsfbox {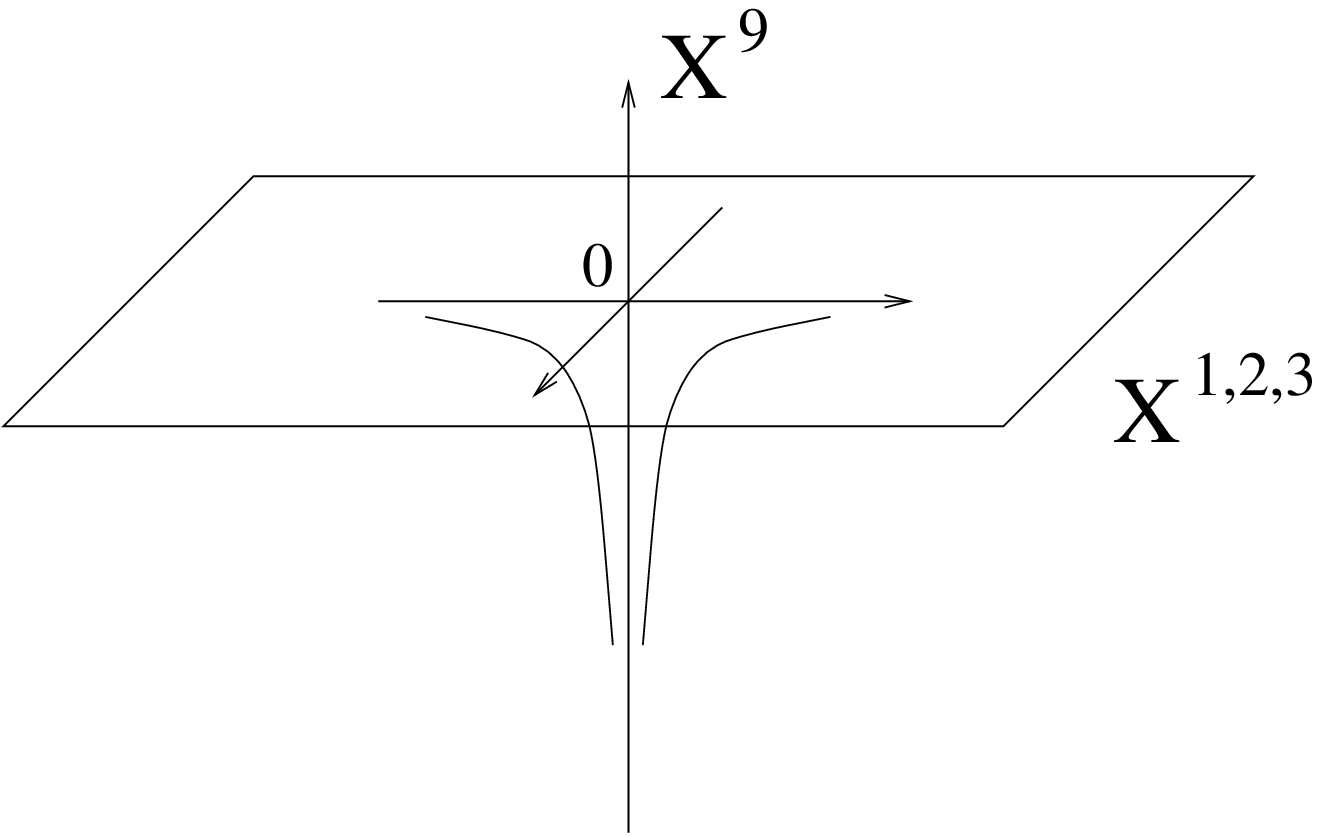}
\end{minipage} \\
 The BI Dyon
\end{center}
$X^{9}$ diverges at the origin, where the electric and magnetic charges sit. The Gauss law and the Bianchi identity are modified to $ \partial_{a} E^{a} = q \delta ({\bf r}) $ , $ \partial_{a} B^{a} = g \delta ({\bf r}) $ .

For this configuration, non-vanishing central charges are
\begin{eqnarray}
Z_{1}^{9} & = & -2 e^{-\phi} g X^{9}(0) \nonumber \\
Z_{3}^{9} & = & 2 q X^{9}(0) \nonumber \\
Z_2^{1 2 3} / i & = & 2 e^{-\phi} \int d^{3} \sigma . 
\label{cha-p3}
\end{eqnarray}
The Bogomol'nyi inequality is
\begin{eqnarray}
0 & \leq & [ (4P^{0})^{2} - \{ (Z_{2}^{1 2 3}/i + \sqrt{(Z_{1}^{9})^{2}+(Z_{3}^{9})^{2}} \}^{2} ]^{8} \nonumber \\
 & & [ (4P^{0})^{2} - \{ (Z_{2}^{1 2 3}/i - \sqrt{(Z_{1}^{9})^{2}+(Z_{3}^{9})^{2}} \}^{2} ]^{8} .  
\label{ineq-p3}
\end{eqnarray}
The configuration saturates this inequality  
\begin{eqnarray}
P^{0} & = & \frac{1}{2} \{ Z_{2}^{1 2 3}/i + \sqrt{(Z_{1}^{9})^{2}+(Z_{3}^{9})^{2}} \} \nonumber \\
 & = & e^{-\phi} \int d^{3} \sigma + \sqrt{q^{2}+e^{-2\phi}g^{2}} (-X^{9}(0)) 
\label{ene-p3}
\end{eqnarray}
The first term is the mass of the D-3-brane and the second term is the mass of the $(q,g)$-string, so this represents $\frac{1}{4}$ supersymmetric marginally BPS state.  

Again we return to the case where axion background exists. The action is \cite{APPS}
\begin{eqnarray}
S = \int d^{4} \sigma {\cal L}_{DBI} + \int L_{WZ} + \int \frac{1}{2} C_{0} F F .
\label{action-p3_da}
\end{eqnarray}
${\cal L}_{DBI}$ and $L_{WZ}$ are same as (\ref{DBI}) and (\ref{WZ}) . 
The form of $Z_{3}^{m}$ is changed 
\begin{eqnarray}
Z_{3}^{m} = -2 \int d^{3} \sigma \partial_{a} X^{m} E^{a} + 2C_{0} \int d^{3} \sigma \partial_{a} X^{m} B^{a} , 
\label{Z3_da}
\end{eqnarray}
while $Z_{1}^{m}$ and $Z_{2}^{m_{1} m_{2} m_{3}}$ are unchanged. 

For this background, the BI Dyon solution is ( Appendix B ) 
\begin{eqnarray}
E^{a} & = & \frac{q}{4 \pi r^{2}} \hat{r}^{a} \nonumber \\
B^{a} & = & \frac{g}{4 \pi r^{2}} \hat{r}^{a} \nonumber \\
X^{9} & = & - \frac{\sqrt{e^{2\phi}(q-C_{0}g)^{2}+g^{2}}}{4 \pi r} 
\label{sol-p3_da}
\end{eqnarray}
Non-vanishing components of the central charges are 
\begin{eqnarray}
Z_{3}^{9} & = & 2 ( q - C_{0} g ) X^{9}(0) \nonumber \\
Z_{1}^{9} & = & -2 e^{-\phi} g X^{9}(0) \nonumber \\
Z_{2}^{1 2 3} / i & = & 2 e^{-\phi} \sigma \int d^{3} \sigma .
\label{cha-p3_da}
\end{eqnarray}
From (\ref{ineq-p3}),(\ref{cha-p3_da})  
\begin{equation}
P^{0} \geq e^{-\phi} \int d^{3} \sigma + \sqrt{ ( q - C_{0} g )^{2} + e^{ -2 \phi} g^{2} } (-X^{9}(0)) . 
\label{ineq-p3_da}
\end{equation}
Again, the second term is the mass of $(q,g)$-string. The inequality is saturated by the solution (\ref{sol-p3_da}) , so this is $\frac{1}{4}$ supersymmetric marginally BPS state. 

In this letter, we investigated BPS properties of the charged solitons of the D-brane worldvolume theories, from the viewpoint of the target-space supersymmetry. Our results agree with those obtained previously. We also extended our BPS analysis to the case where axion background is turned on. 

In our analysis, we used the Poisson bracket algebra of the supersymmetry charges. Of course, the classical counter-part of the (anti-)commutator is the Dirac bracket, so our analysis is not complete. Here we assumed that the forms of the central charges are unchanged when they are calculated by using the Dirac bracket. 

Finally, we comment on future analysis. The 3-string junction as a solution of the D-string worldsheet theory is not symmetric, in the sense that one of the strings is necessarily a F-string ( the end of which is the electric charge on D-string ) . Recently, the $SL(2,{\bf Z})$-covariant string action was proposed \cite{T,CT} . It will be interesting to construct the (charged) soliton solutions of this theory, because we expect to obtain symmetric 3-string junction solution \footnote{ In \cite{M} , the $SL(2,{\bf Z})$-covariant string theory with the source terms of the worldsheet gauge fields was considered, and the configuration, which represents the 3-string junction, was found. } . In the theory of \cite{T,CT},the tension is a dynamical variable. Theories of this type for general $p>1$ were also constructed \cite{CW,BT_re} . It will be also interesting to find their soliton solutions. To examine their BPS properties in the way we took in this article, we need to calculate the Poisson bracket algebra of the supersymmetry charges. The results will be reported elsewhere \cite{watasi}. 

\begin{flushleft}
{\large Acknowledgements}
\end{flushleft}

The author would like to thank Prof. R. Nakayama for critical reading of the manuscript and advice. The author also thank K. Hayasaka for many discussions and Prof. K. Ishikawa for encouragement. 

\begin{flushleft}
{\large Appendix A}
\end{flushleft}

In type IIB theory, we can include non-vanishing axion background \cite{APPS} . The action can be written collectively for any odd $p$ 
\begin{eqnarray}
S & = & \int d^{p+1} \sigma {\cal L}_{DBI} + \int L_{WZ} + \int L_{F} \nonumber \\ 
L_{F} & = & \int C_{0} e^{-F} . 
\label{action_da}
\end{eqnarray}  

For any variation of $A$ , the variation of $L_{F}$ is 
\begin{equation}
\delta L_{F} = \int d [ - C_{0} \delta A e^{-F} ] - \int C_{0} \delta A dF e^{-F} .
\label{del-Lf}
\end{equation} 
For singular configuration of $A$ , $dF$ is non-vanishing, so the second term could be non-zero. However, for the supersymmetry transformation, if we set the configuration bosonic ($\theta=0$) , the second term vanishes. So, for bosonic configuration, the action (\ref{action_da}) is invariant up to total derivative under the supersymmetry transformation as (\ref{susy}) . The corresponding supersymmetry charge is 
\begin{equation}
Q \epsilon = Q_{1} \epsilon + Q_{2} \epsilon + Q_{3} \epsilon 
\label{Q_da}
\end{equation}
where
\begin{equation}
Q_{3} \epsilon = C_{0} \int d^{p} \sigma [ \delta_{\epsilon}A e^{-F} ]_{\bf p} 
\label{Q_F}
\end{equation}
and $Q_{1}$ , $Q_{2}$ are same as (\ref{Q}) . 

The supersymmetry algebra is calculated as 
\begin{eqnarray}
\{ Q_{\alpha A} , Q_{\beta B} \} & = & -2 (C \Gamma_{m})_{\alpha \beta} \delta_{A B} \int d^{p} \sigma p^{m} \nonumber \\
 & & -2 (\tau_{3})_{A B} (C \Gamma_{m})_{\alpha \beta} \int d^{p} \sigma \partial_{a} X^{m} E^{a} \nonumber \\
 & & - (\tilde{\tau}_{J})_{A B} (C \Gamma_{M})_{\alpha \beta} \frac{2e^{-\phi}}{2^{5}} (1 + (-)^{\sigma_{M} + J}) \int d^{p} \sigma [U_{J}^{M}]_{\bf p} \nonumber \\
 & & - 2 C_{0} (C \Gamma_{m})_{\alpha \beta} (\tau_{3})_{A B} \int [ d X^{m} e^{-F} ]_{\bf p} . 
\label{algebra_da}
\end{eqnarray}
As mentioned above, we set the configuration bosonic \footnote{ It should be noted that, when we calculate the algebra, we must keep the terms up to the first order in $\theta$ in the supersymmetry charge (\ref{Q_da}) . } . The fourth term comes from $ \{ Q_{1} , Q_{3} \} $ and $ \{ Q_{3} , Q_{1} \} $ , which changes the form of $ Z_{3}^{m} $ . Note that $Q_{2}$ and $Q_{3}$ contain no conjugate momentum. From this, we see immediately (\ref{Z1_da}) , (\ref{Z3_da}) . 

\begin{flushleft}
{\large Appendix B}
\end{flushleft}

In this appendix, we derive the 3-string junction solution for $p=1$ and the BI Dyon solution for $p=3$ , in the presence of axion background. To find the solutions, we use a method based on Hamiltonian formalism, which is same as the one in \cite{GGT} . 

We are interested in bosonic configuration, so we consider bosonic theory from the beginning. The bosonic action is 
\begin{equation}
S^{boson} = - \int d^{p+1} \sigma e^{-\phi} \sqrt{-det(G_{\mu \nu}+F_{\mu \nu})} + \int C_{0} e^{-F} . 
\label{action_F}
\end{equation}
The constraints are 
\begin{eqnarray}
T_{a} & = & p_{m} \partial_{a} X^{m} + \tilde{E}^{b} F_{a b} \approx 0 \nonumber \\
H & = & \frac{1}{2} \{ p^{2} + \tilde{E}^{a} G_{a b} \tilde{E}^{b} + e^{-2\phi} det(G_{a b}+F_{a b}) \} \approx 0 
\label{constraint1}
\end{eqnarray}
where $ \tilde{E}^{a} = E^{a} - \frac{\partial {\cal L}_{F}}{\partial F_{0 a}} $ . If we replace ${\cal L}_{F}$ with ${\cal L}_{WZ}$ , these are quite similar to the bosonic constraints in \cite{HK1,HK2} . 

Now we take the static gauge $X^{\mu}=\sigma^{\mu}$ , moreover configurations are assumed to be static, that is, $\dot{X}^{i}=0$ and $p_{i}=0$ ( $\cdot$ denotes time derivative ) . So (\ref{constraint1}) reduces to 
\begin{equation}
(p^{0})^{2} = \tilde{E}^{c} \tilde{E}^{d} F_{a c} F_{b d} \delta^{a b} + \tilde{E}^{a} \tilde{E}^{b} G_{a b} + e^{-2\phi} det(G_{a b}+F_{a b}) . 
\label{constraint2} 
\end{equation}

For $p=1$ case, where the configuration is automatically pure electric ($F_{a b}=0$) , we have 
\begin{eqnarray}
(p^{0})^{2} & = & (\tilde{E})^{2} \{ 1 + (\partial X^{9})^{2} \} + e^{-2\phi} \{ 1 + (\partial X^{9})^{2} \} \nonumber \\
 & = & (e^{-\phi} \pm \tilde{E} \partial X^{9})^{2} + (\tilde{E} \mp e^{-\phi} \partial X^{9})^{2}  
\label{constraint-p1}
\end{eqnarray} 
where $ \tilde{E} = E + C_{0} $ . Here we assumed that only one scalar field ($X^{9}$) is excited. From this, we see that if $ \tilde{E} = \pm e^{-\phi} \partial X^{9} $ , the energy is minimized. Taking into account the modified Gauss law $ \partial E = q \delta $ , we obtain the solution (\ref{sol-p1_da}) . 

Next we turn to $p=3$ case, where we can have electric-magnetic solution. We rewrite (\ref{constraint2}) as  
\begin{eqnarray}
(p^{0})^{2} & = & (\tilde{E} \times B)^{2} + e^{-2\phi} \{ 1 + (\partial X^{9})^{2} + B^{2} + (B \cdot \partial X^{9})^{2} \} + \tilde{E}^2 + (\tilde{E} \cdot \partial X^{9})^{2} \nonumber \\ 
 & = & (\tilde{E} \times B)^{2} + \{ e^{-\phi} + \sin \omega (\tilde{E} \cdot \partial X^{9}) + \cos \omega e^{-\phi} (B \cdot \partial X^{9}) \}^{2} \nonumber \\
 & & + (\tilde{E} - \sin \omega e^{-\phi} \partial X^{9})^{2} + (e^{-\phi} B - e^{-\phi} \cos \omega \partial X^{9})^{2} \nonumber \\
 & & + \{ \cos \omega (\tilde{E} \cdot \partial X^{9}) - \sin \omega e^{-\phi} (B \cdot \partial X^{9}) \}^{2}     
\label{constraint-p3}
\end{eqnarray}
where $ \tilde{E}_{a} = E_{a} - C_{0} B_{a} $ . The energy is minimum, when $ \tilde{E}_{a} = \sin \omega e^{-\phi} \partial_{a} X^{9} $ and $ B_{a} = \cos \omega \partial_{a} X^{9} $ . Together with the modified Gauss law $ \partial_{a} E^{a} = q \delta $ and the modified Bianchi identity $ \partial_{a} B^{a} = g \delta $ , (\ref{constraint-p3}) leads to the solution (\ref{sol-p3_da}) .

\end{document}